\documentclass[aps,prb,twocolumn,superscriptaddress,longbibliography]{revtex4-2}

\usepackage{amsmath,amssymb,graphicx,bm}

\usepackage{graphicx}
\usepackage{bm}
\usepackage{color}
\usepackage{xcolor}
\usepackage{epstopdf}
\usepackage{amsmath}
\usepackage{url}
\usepackage{braket}
\usepackage{esvect}
\definecolor{darkGreen}{RGB}{0,110,0}
\definecolor{darkBlue}{RGB}{0,0,130}
\usepackage[colorlinks=true, urlcolor=darkBlue,citecolor=darkGreen,linkcolor=darkBlue]{hyperref}

\begin{document}
	
	\title{Current-driven reduction of topological protection in multichannel superconductors}
	
	\author{Alfonso Maiellaro}
	\affiliation{CNR-SPIN, c/o Università di Salerno, IT-84084 Fisciano (SA), Italy}
	\affiliation{Dipartimento di Fisica "E.R. Caianiello", Università di Salerno, Via Giovanni Paolo II, 132, I-84084 Fisciano (SA), Italy}
	
	\date{\today}
	
	\begin{abstract}
		We investigate the robustness of topological phases in a Kitaev ladder composed of two coupled superconducting chains under the perturbing influence of a finite charge current. By introducing an effective Hamiltonian depending on the quasiparticle momentum induced by the current, we show that the two-mode topological phase, present in the isolated ladder, is fragile against a finite current flux. 
		To characterize this behavior, we combine bulk topological invariants with real-space diagnostics, including the edge--edge quantum conditional mutual information $I_{ee}$, which provides an entanglement-based signature of topological order. We further distinguish the gapped class-D sectors from a current-induced gapless superconducting regime, which emerges when the upper BdG branch crosses zero at generic momenta and the standard gapped topological invariant is no longer applicable. Our results provide an effective framework to describe how current injection and measurement processes can affect topological protection in superconducting nanostructures.
	\end{abstract}
	
	\maketitle
	\section{Introduction}
	Topological superconductivity has attracted a steadily growing interest in the last two decades, mainly due to its potential applications in quantum technologies~\cite{PhysRevLett.100.096407,PhysRevLett.105.177002,PhysRevLett.105.077001,PhysRevB.84.144526}. In particular, one-dimensional $p$-wave superconductors host Majorana bound states (MBSs), zero-energy excitations localized at the system edges, which obey non-Abelian statistics and are regarded as key ingredients for fault-tolerant quantum computation~\cite{pachos2012}.\\
	Among the simplest and most paradigmatic models supporting topological superconductivity, there is the Kitaev chain~\cite{Kitaev_2001}. Despite its minimal structure, this model captures the essential physics of Majorana modes and provides a powerful framework to investigate the robustness of topological phases against perturbations, disorder, and geometrical extensions~\cite{PhysRevB.88.064506,PhysRevB.94.115166,Maiellaro2020,Neven_2013,PhysRevB.97.041109,PhysRevB.88.165111,Maiellaro2021,PhysRevLett.113.156402,PhysRevB.95.195160,condmat9010020}. In this context, ladder geometries, consisting of coupled Kitaev chains, represent a natural generalization that allows for the emergence of richer topological phases, including multimode regimes characterized by more than a single Majorana mode per edge~\cite{PhysRevB.89.174514,Maiellaro_2019,PhysRevLett.105.227003,Maiellaro2018,PhysRevB.96.035306,condmat6020015,condmat7010026}.\\
	In equilibrium conditions, topological phases of mean-field Hamiltonians are naturally described by bulk topological invariants within the Altland--Zirnbauer tenfold classification scheme~\cite{PhysRevB.55.1142}. The ten symmetry classes allow one to identify the appropriate topological invariant for a given bulk Hamiltonian according to the system dimensionality and to the simultaneous presence or absence of particle-hole, time-reversal, and chiral symmetries. These invariants capture the topology of the bulk band structure, provide the corresponding phase diagrams in the thermodynamic limit, and identify the gap-closing points associated with topological phase transitions.\\
	Despite the rich literature on closed topological systems hosting MBSs, a crucial issue concerns their stability under realistic operating conditions. In most experimental setups, the detection of Majorana modes relies on transport measurements, such as tunneling spectroscopy~\cite{PhysRevLett.98.237002,PhysRevLett.103.237001,PhysRevLett.119.136803,PhysRevLett.110.126406}  or Josephson effects~\cite{PhysRevB.84.180502,PhysRevB.107.L201405,PhysRevB.103.144502,GUARCELLO2024115596}, which necessarily involve the injection of a current into the system. This introduces genuine nonequilibrium perturbations that may affect the topological properties of the system. Therefore, understanding how topological phases respond to current-induced effects is of fundamental importance, both from a conceptual and an application-oriented perspective.\\
	In this context, several studies have investigated how coupling to external reservoirs, dissipation, and nonequilibrium conditions can modify the topology of superconducting systems and alter the bulk--edge correspondence~\cite{PhysRevB.104.134516,PhysRevB.101.094502}. These observations also motivate the search for real-space quantities capable of characterizing topological phases beyond the standard momentum-space classification, especially in finite-size systems or in the presence of measurement-induced perturbations. Entanglement-based observables~\cite{z4z8,PhysRevB.110.L220410}, such as the edge-edge quantum conditional mutual information~\cite{Maiellaro2023,PhysRevB.106.155407,PhysRevResearch.4.033088,PhysRevB.107.115160}, provide a particularly suitable framework to probe the nonlocal correlations associated with topological edge modes.\\
	Despite these efforts, a full nonequilibrium treatment of current-driven topological superconductors, however, remains challenging. For this reason, an effective and physically transparent strategy has been introduced for a single Kitaev chain~\cite{Maiellaro2023}, consisting in the incorporation of a finite quasiparticle momentum $q$ in the hopping amplitudes. In this framework, the hopping phase factor $e^{iq}$ mimics the presence of a charge current, shifts the band dispersion and, in the presence of $p$-wave superconducting correlations, induces a finite momentum $2q$ of the Cooper pairs~\cite{Maiellaro2023}, parallel to the current direction. This approach, analogous to a Peierls substitution, captures the essential effects of current flow within a static Hamiltonian description. Already in Ref.~\cite{Maiellaro2023}, the single-chain spectrum was shown to develop generic-momentum zero-energy crossings above the critical quasiparticle momentum $q_c=\arcsin(\Delta/t)$, beyond which the gapped Majorana phase is lost. The same band-overlap regime was subsequently discussed in Ref.~\cite{PhysRevResearch.6.033060} as a gapless superconducting phase associated with a Lifshitz transition and with the appearance of unpaired particle and hole pockets. However, in a single Kitaev chain the topological invariant is restricted to two values, limiting the richness of possible phases and thus the impact of $q$.\\  
	In this work, we extend this effective description to a Kitaev ladder~\cite{Maiellaro2018} and investigate the interplay between current-induced symmetry breaking and topological classification. In particular, we show that a finite $q$ breaks time-reversal symmetry and, consequently, chiral symmetry, driving the system from the BDI class, characterized by an integer ($\mathbb{Z}$) invariant, to the D class, where the gapped sectors are characterized by a $\mathbb{Z}_2$ invariant. This symmetry reduction has direct consequences in ladder systems, where multimode topological phases are allowed in equilibrium. In particular, the phase supporting two Majorana modes per edge is suppressed by the current, providing a clear signature of the reduced robustness of higher topological sectors under measurement-induced perturbations.\\
	The ladder geometry also allows us to explore regimes where additional symmetries emerge, leading to nontrivial deviations from the D class. In particular, under a current acting along the chains, with no additional flux or superconducting phase gradient across the transverse direction, we identify conditions under which the system effectively decomposes into independent channels, preserving higher topological sectors even in the presence of a finite current.\\
	To characterize the topological phases, we combine bulk and real-space approaches. On the one hand, we first identify possible gapless regions from the bulk spectrum and then determine the topology of the remaining gapped sectors from the appropriate bulk invariant.On  the other hand, we employ real-space diagnostics, including the edge-edge quantum conditional mutual information~\cite{PhysRevResearch.4.033088,PhysRevB.107.115160}, which quantifies the nonlocal correlations between the system edges and provides a robust entanglement-based signature of topological order.\\
	The paper is organized as follows. In Sec.~\ref{Sec2}, we introduce the model and discuss its symmetry properties and bulk topological invariant. In Sec.~\ref{Sec3}, we present the numerical results, including the phase diagrams and the analysis of the bulk--edge correspondence. In Sec.~\ref{Sec4}, we discuss the implications of our findings. The distinction between gapped and gapless superconducting regimes of the general ladder is discussed in Appendix~\ref{App0}. A gauge-equivalent representation of the real-space Hamiltonian, which makes explicit the finite momentum $2q$ acquired by the Cooper pairs, is presented in Appendix~\ref{AppA}. The generalized exchange-symmetric limit $t_1=0$ is discussed in Appendix~\ref{AppB}.
	\begin{figure*}
		\centering
		\includegraphics[scale=0.2]{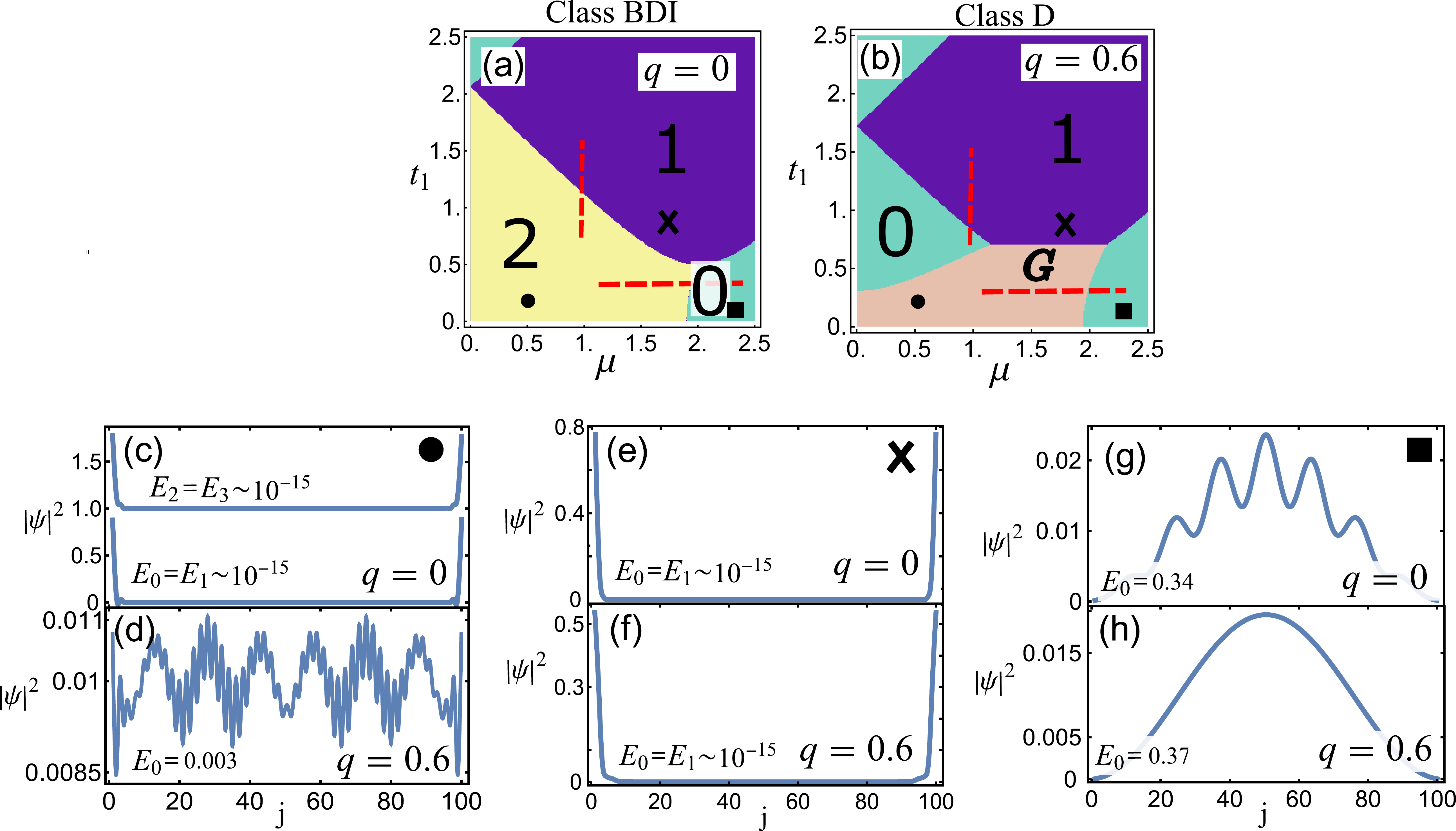}
		\caption{(a),(b): phase diagrams of the Kitaev ladder for $\Delta_1=0.5$, evaluated at $q=0$ and $q=0.6$, respectively. The diagrams are constructed using the label $\mathcal{N}$. In panel (a), the labels $0$, $1$, and $2$ denote the number of Majorana modes per edge in the gapped BDI phases. In panel (b), $0$ and $1$ denote the trivial and topological gapped class-D sectors, respectively, whereas $G$ identifies the gapless superconducting region defined by the band-overlap criterion discussed in Sec.~\ref{Sec2} and Appendix~\ref{App0}. Panels (c)--(h) display the probability densities of the lowest-energy BdG modes for representative points selected from the phase diagrams. The real-space calculations are performed on a ladder composed of two chains, each of length $L=100$. The corresponding BdG energies of the modes are reported in the insets.}
		\label{Figure0}
	\end{figure*}
	\section{Model and symmetry analysis}
	\label{Sec2}
\begin{figure*}
		\centering
		\includegraphics[scale=0.4]{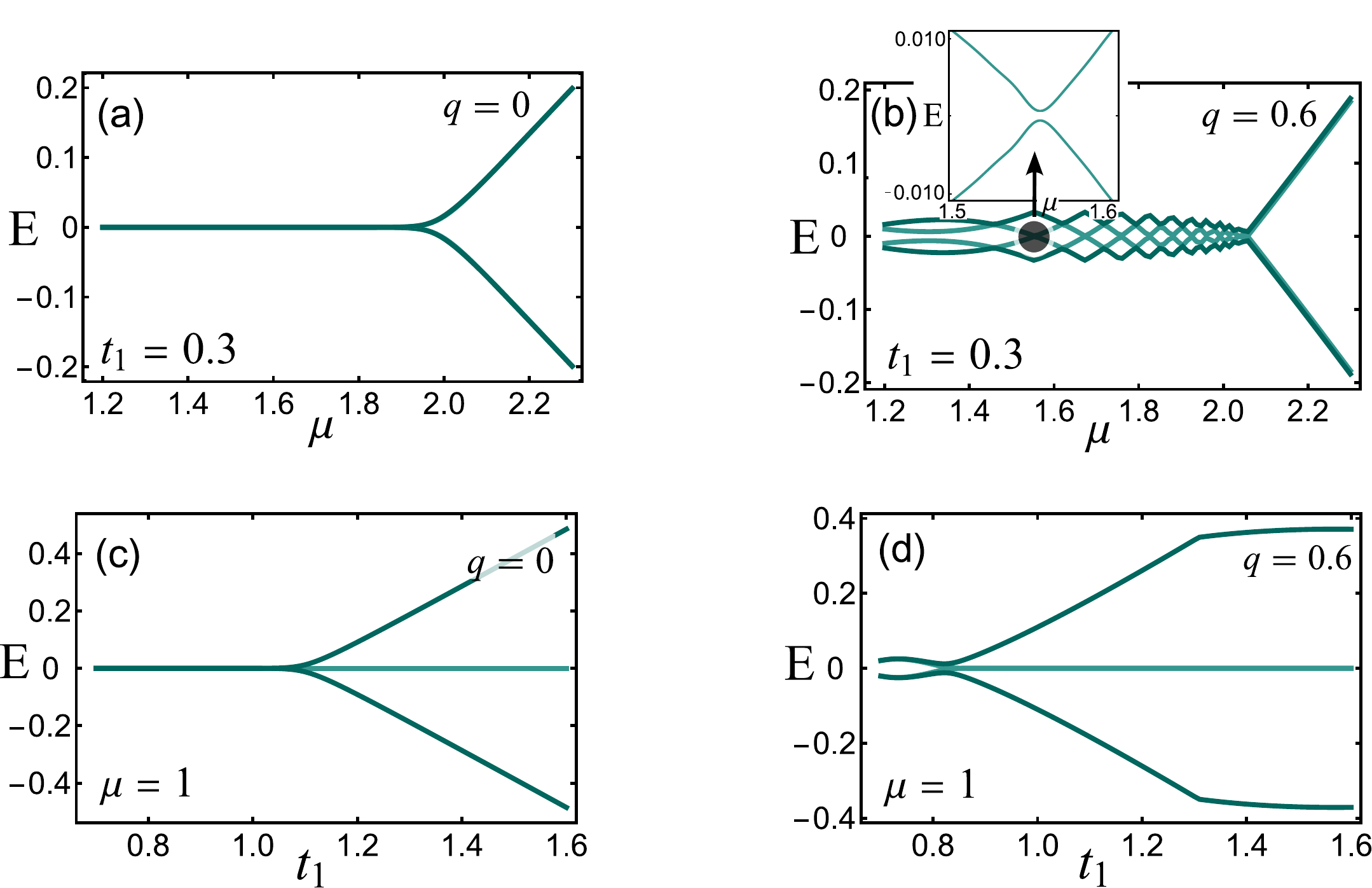}
		\caption{Panels (a) and (b) show horizontal cuts of the phase diagrams along the red lines indicated in Fig.~\ref{Figure0}(a) and Fig.~\ref{Figure0}(b), respectively, while panels (c) and (d) show the corresponding vertical cuts. In all panels, the four BdG eigenvalues closest to zero are plotted for a ladder composed of two chains, each of length $L=100$. The spectra are shown both in the absence of current, $q=0$, in panels (a) and (c), and in the current-carrying regime, $q=0.6$, in panels (b) and (d). The interchain pairing is fixed to $\Delta_1=0.5$ in all panels. The accumulation of low-energy bulk levels in the portion of panel (b) crossing the $G$ region is the finite-size counterpart of the gapless bulk spectrum.}
		\label{Figure1}
	\end{figure*}
	The Kitaev ladder introduced in Ref.~\cite{Maiellaro2018,nano9060894}, consists of two spinless $p$-wave superconducting chains coupled by both interchain hopping $t_1$ and interchain pairing $\Delta_1$ terms, under the perturbing influence of a finite charge current. Following Ref.~\cite{Maiellaro2023}, the current is modeled by introducing a finite quasiparticle momentum $q$ in the longitudinal hopping amplitudes. The resulting real-space Hamiltonian reads:
	\begin{eqnarray}
		\nonumber
		H &=& \sum_{j,l}
		\Big[
		- \mu c^\dagger_{j,l} c_{j,l}
		- t e^{i q} c^\dagger_{j,l} c_{j+1,l}
		+ \Delta c_{j,l} c_{j+1,l}
		+ \mathrm{h.c.}
		\Big] \\
		&&+ \sum_j
		\Big[
		- t_1 c^\dagger_{j,1} c_{j,2}
		+ \Delta_1 c_{j,1} c_{j,2}
		+ \mathrm{h.c.}
		\Big].
		\label{eq:H_real}
	\end{eqnarray}
	Here $l=1,2$ labels the two chains, $\mu$ is the chemical potential, $t$ the hopping amplitude, and $\Delta$ the longitudinal $p$-wave pairing. Throughout this work we take $t_1$ and $\Delta_1$ to be real. This choice corresponds to a ladder without a transverse magnetic flux or externally imposed superconducting phase difference between the two legs. The phase factor $e^{iq}$ effectively accounts for a finite quasiparticle momentum induced by a charge current, shifting the band dispersion as $k \to k - q$~\cite{Maiellaro2023}. An equivalent gauge representation, presented in Appendix~\ref{AppA}, makes explicit that the Cooper pairs acquire a finite momentum $2q$, thereby showing that both the intraleg pairing $\Delta$ and the interleg pairing $\Delta_1$ contribute to the current-carrying condensate.\\
	When $q=0$, the Hamiltonian satisfies particle-hole ($\mathcal{P}$), time-reversal ($\mathcal{T}$), and chiral ($\mathcal{C}$) symmetries. As a consequence, the system belongs to the BDI class of the Altland--Zirnbauer classification~\cite{PhysRevB.55.1142}, characterized by an integer ($\mathbb{Z}$) topological invariant and allowing for multiple Majorana modes per edge~\cite{Maiellaro2018}.\\
	The situation changes when a finite current is introduced. The phase $q$ explicitly breaks time-reversal symmetry by selecting a preferred propagation direction for quasiparticles. Since chiral symmetry is defined as $\mathcal{C}=\mathcal{P}\mathcal{T}$, the latter symmetry is also broken. The system symmetry class is therefore reduced to class D, where only particle-hole symmetry survives. Whenever the bulk spectrum remains gapped, the corresponding topological classification is $\mathbb{Z}_2$ and at most a single protected Majorana mode per edge is expected. A finite $q$, however, may also produce an indirect gap closing at generic momentum, leading to a gapless superconducting phase for which the standard gapped tenfold invariant is not defined. From the homotopy-theoretic viewpoint, the above symmetry classification can be formulated in terms of the classifying spaces associated with the corresponding Clifford-extension problems. For a one-dimensional gapped free-fermion Hamiltonian, the stable topological classification is determined by $\pi_0(R_{z-d})$, where $z$ labels the real Altland--Zirnbauer class and $d=1$ is the spatial dimension~\cite{Kitaev2009}. In the absence of current, the ladder belongs to class BDI, corresponding to
\begin{equation}
    \pi_0(R_{1-1})=\pi_0(R_0)=\mathbb{Z},
\end{equation}
which accounts for the integer winding number and for the possible presence of
multiple Majorana modes at each edge. When a finite current breaks
time-reversal and chiral symmetries, the system is reduced to class D, for
which
\begin{equation}
    \pi_0(R_{2-1})=\pi_0(R_1)=\mathbb{Z}_2.
\end{equation}
For a gapped class-D spectrum, the Pfaffian invariant provides an explicit realization of the two disconnected components of the class-D space. More generally, topological charges can be interpreted as homotopy classes of maps into the corresponding classifying space~\cite{Kitaev2009,Bzduszek2017}.\\
These observations can be made explicit by introducing the Nambu spinor $\Psi_k = (c_{k,1}, c^\dagger_{-k,1}, c_{k,2}, c^\dagger_{-k,2})^T$, which allows the Hamiltonian to be written in Bogoliubov--de Gennes form in momentum space:
	\begin{equation}
		H = \frac{1}{2} \sum_k \Psi_k^\dagger H(k) \Psi_k,
	\end{equation}
	where
	\begin{equation}
		H(k)=
		\begin{pmatrix}
			-\epsilon_k(q) & \Delta_k & t_1 & -\Delta_1 \\
			-\Delta_k & \epsilon_k(-q) & \Delta_1 & -t_1 \\
			t_1 & \Delta_1 & -\epsilon_k(q) & \Delta_k \\
			-\Delta_1 & -t_1 & -\Delta_k & \epsilon_k(-q)
		\end{pmatrix},
		\label{eq:Hk}
	\end{equation}
with $\epsilon_k(q)=2t\cos(k-q)+\mu$ and $\Delta_k=2i\Delta\sin k$.\\
It is straightforward to verify that, for $q=0$, the Hamiltonian fulfills the relations $\mathcal{P}H(k)\mathcal{P}^\dagger=-H(-k)$, $\mathcal{T}H(k)\mathcal{T}^\dagger=H(-k)$, and $\mathcal{C}H(k)\mathcal{C}^\dagger=-H(k)$, with $\mathcal{P}=\mathbb{I}_2\otimes\sigma_x K$, $\mathcal{T}=K$, and $\mathcal{C}=\mathbb{I}_2\otimes\sigma_x$, where $K$ denotes complex conjugation. For $q\neq 0$, in presence of only $\mathcal{P}$, the appropriate bulk topological invariant is given by the Pfaffian of the antisymmetric Hamiltonian written in the Majorana basis:
	\begin{equation}
		U_M = \mathbb{I}_2 \otimes 
		\frac{1}{\sqrt{2}}
		\begin{pmatrix}
			1 & 1 \\
			i & -i
		\end{pmatrix}.
	\end{equation}
	
	The $\mathbb{Z}_2$ invariant is then
	\begin{equation}
		\nu = \mathrm{sgn}\big[\mathrm{Pf}(iH_M(0))\,\mathrm{Pf}(iH_M(\pi))\big],
		\label{TI1}
	\end{equation}
	with $\mathrm{Pf}(iH_M(0/\pi)) = (\mu \pm 2t\cos q)^2 + \Delta_1^2 - t_1^2$.\\
Equation~\eqref{TI1} is therefore used only after verifying that the bulk spectrum is gapped. At finite $q$, the term proportional to the identity in the BdG spectrum changes the occupation of the quasiparticle bands and may drive the upper branch below zero energy at generic momentum, producing the gapless superconducting regime discussed for the current-carrying Kitaev chain in Refs.~\cite{Maiellaro2023,PhysRevResearch.6.033060}. For the general ladder we identify this regime through the condition
\begin{equation}
\min_{k\in[-\pi,\pi]} E_{+}(k)<0,
\label{eq:gaplesscriterion}
\end{equation}
where $E_{+}(k)$ is the inner upper BdG branch given explicitly in Appendix~\ref{App0}. For the finite-current phase diagrams we therefore introduce the following phase label, which supplements---rather than replaces---the gapped topological invariant:
\begin{equation}
\mathcal{N}=\begin{cases}
G, & \min_k E_+(k)<0,\\[2pt]
(1-\nu)/2, & \min_k E_+(k)>0.
\end{cases}
\label{eq:phaseclassifier}
\end{equation}
Thus, $\mathcal{N}=1$ and $0$ denote the topological and trivial gapped class-D sectors, respectively, while $G$ denotes the gapless superconducting phase.\\The condition $\min_{k\in(0,\pi)}E_+(k)=0$ marks the onset of the gapless superconducting regime through a zero-energy crossing at generic momentum. This condition must be distinguished from the gap closings at the particle-hole-invariant momenta $k=0,\pi$, which determine the sign changes of the Pfaffian $\mathbb{Z}_2$ invariant.\\
Interestingly, the ladder may possess additional discrete exchange symmetries whenever at least one of the three couplings $t_1$, $\Delta_1$, or $\Delta$ vanishes. In each of these special limits, a suitable generalized bonding/antibonding transformation block-diagonalizes the BdG Hamiltonian into two symmetry sectors. The explicit form of the transformation depends on which coupling is set to zero.\\
        \begin{figure*}
		\centering
		\includegraphics[scale=0.8]{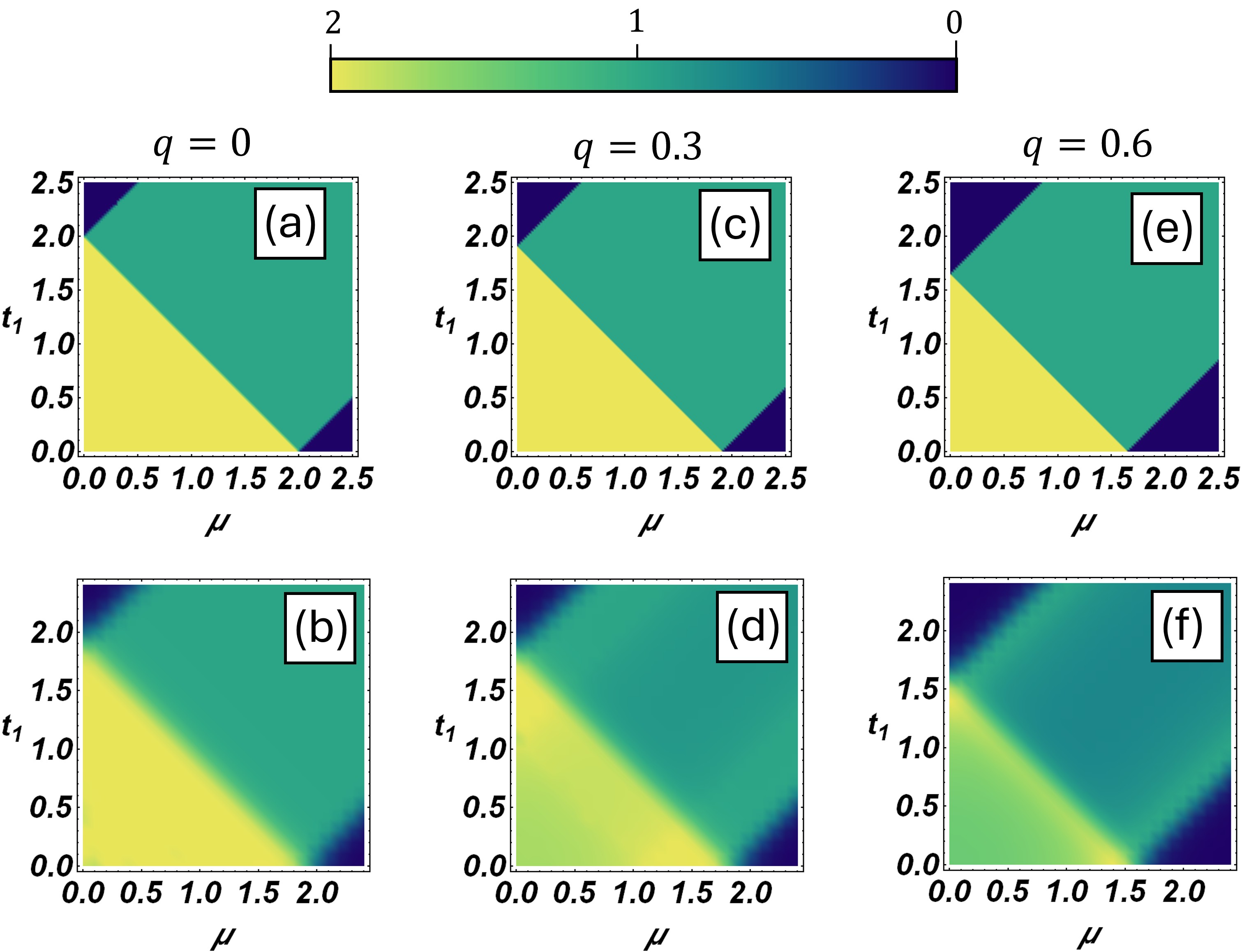}
		\caption{Panels (a), (c), and (e): phase diagrams for the symmetry-protected limit $\Delta_1=0$, constructed from the quantity $Q=Q_{+}+Q_{-}$ introduced in Sec.~II, for increasing values of the current flux $q=0$, $0.3$, and $0.6$, respectively. Panels (b), (d), and (f): corresponding phase diagrams obtained from the edge--edge quantum conditional mutual information $I_{ee}$. The agreement between the two sets of panels highlights the consistency between the momentum-space and real-space entanglement-based characterizations. The entanglement calculations are performed on a ladder composed of two chains, each of length $L=50$, with edge sizes fixed to $L_A=L_B=12$ sites per chain.}
		\label{Figure2}
	\end{figure*}
When $\Delta_1=0$, $t_1 \neq 0$ and $\Delta \neq 0$ the additional symmetry corresponds to the ordinary exchange of the two legs of the ladder, $c_{j,1}\leftrightarrow c_{j,2}$, which is described in momentum space by the operator $\mathcal{M}=\sigma_x\otimes\mathbb{I}_2$. In this case one verifies that $[H(k),\mathcal{M}]=0$, so that the Hamiltonian can be decomposed into bonding and antibonding sectors. This additional symmetry has nontrivial consequences for the topological classification, since it protects a two-mode topological phase even in the presence of a finite current.\\
These become clear upon performing a unitary transformation to the bonding/antibonding basis, $c_{j,\pm} = (c_{j,1} \pm c_{j,2})/\sqrt{2}$, which leads to the following representation:
	\begin{equation}
		H'(k)=
		\begin{pmatrix}
			-\epsilon_k^-(q) & \Delta_k & 0 & \Delta_1 \\
			-\Delta_k & \epsilon_k^-(-q) & -\Delta_1 & 0 \\
			0 & -\Delta_1 & -\epsilon_k^+(q) & \Delta_k \\
			\Delta_1 & 0 & -\Delta_k & \epsilon_k^+(-q)
		\end{pmatrix},
		\label{eq:Hprime}
	\end{equation}
	with $\epsilon_k^\pm(q)=2t\cos(k-q)+\mu_\pm$ and $\mu_\pm=\mu \pm t_1$.\\
	For $\Delta_1=0$, Eq.~\eqref{eq:Hprime} becomes block diagonal, describing two completely decoupled Kitaev chains with effective chemical potentials $\mu_\pm$. Therefore, despite the breaking of time-reversal symmetry by the current, the ladder effectively decomposes into two independent class-D systems. Each channel is characterized, as long as it remains gapped, by the same $\mathbb{Z}_2$ invariant as the current-carrying Kitaev chain studied in Ref.~\cite{Maiellaro2023}, evaluated at the shifted chemical potential $\mu_\pm$. Thus, the ladder can host $0$, $1$, or $2$ Majorana modes per edge depending on whether none, one, or both channels are topological. The onset of the gapless regime can also be determined analytically in this decoupled limit: the appearance of generic-momentum zero-energy crossings occurs at the critical momentum $q_c=\arcsin(\Delta/t)$, as obtained in Refs.~\cite{Maiellaro2023,PhysRevResearch.6.033060}. Accordingly, the topological invariant introduced for the two decoupled channels can be written as $Q=Q_{+}+Q_{-}$, with $Q_{\pm}=\mathrm{sgn}\!\left[\mathrm{sgn}\!\left[(-\mu_{\pm}+2t\cos q)(-\mu_{\pm}-2t\cos q)\right]+\mathrm{sgn}(q-q_c)\right]$. For $q<q_c$, the relevant gap closings occur at the particle-hole-invariant momenta $k=0,\pi$, whereas for $q>q_c$ generic-momentum band crossings appear and the system enters the gapless superconducting regime.
\begin{figure*}
	\centering
	\includegraphics[scale=0.35]{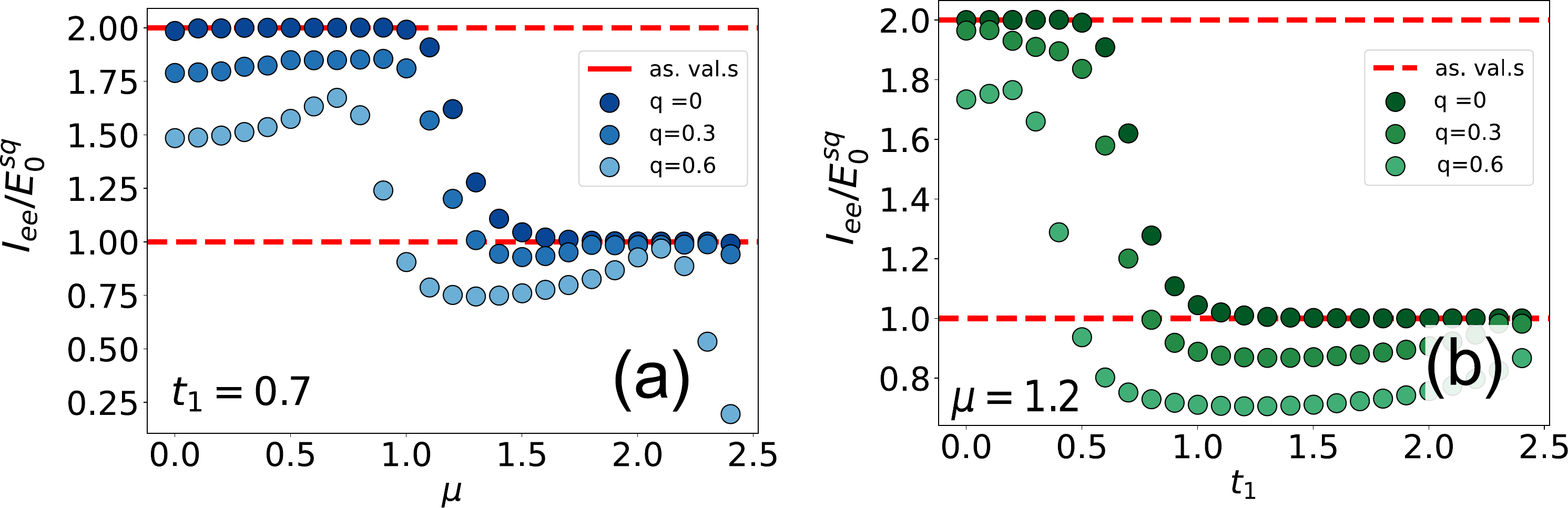}
	\caption{Panel (a) shows a horizontal cut, while panel (b) shows a vertical cut of the phase diagrams obtained from the edge--edge quantum conditional mutual information $I_{ee}$ in Fig.~\ref{Figure2}. The cuts are taken at fixed $t_1=0.7$ in panel (a) and at fixed $\mu=1.2$ in panel (b). The dashed lines indicate the ideal quantized values corresponding to the different topological sectors. Deviations from these quantized plateaus become more pronounced as the current flux $q$ increases, reflecting the growing relevance of finite-size effects.}
	\label{Figure3}
\end{figure*}
In Appendix~\ref{AppB} we show that in the case with $t_1=0$ and $q \neq 0$ the inter-leg pairing $\Delta_1$ acts as a mass term only at the particle-hole-invariant momenta $k=0,\pi$, preventing the topological gap closing there This does not exclude zero-energy crossings at generic momenta: depending on the remaining parameters, this limit is either a trivial gapped superconductor or a gapless superconductor. Therefore, such limit does not support a non-trivial gapped topological phase with Majorana edge states.
\section{Numerical results}
	\label{Sec3}
\begin{figure}
		\centering
		\includegraphics[scale=0.26]{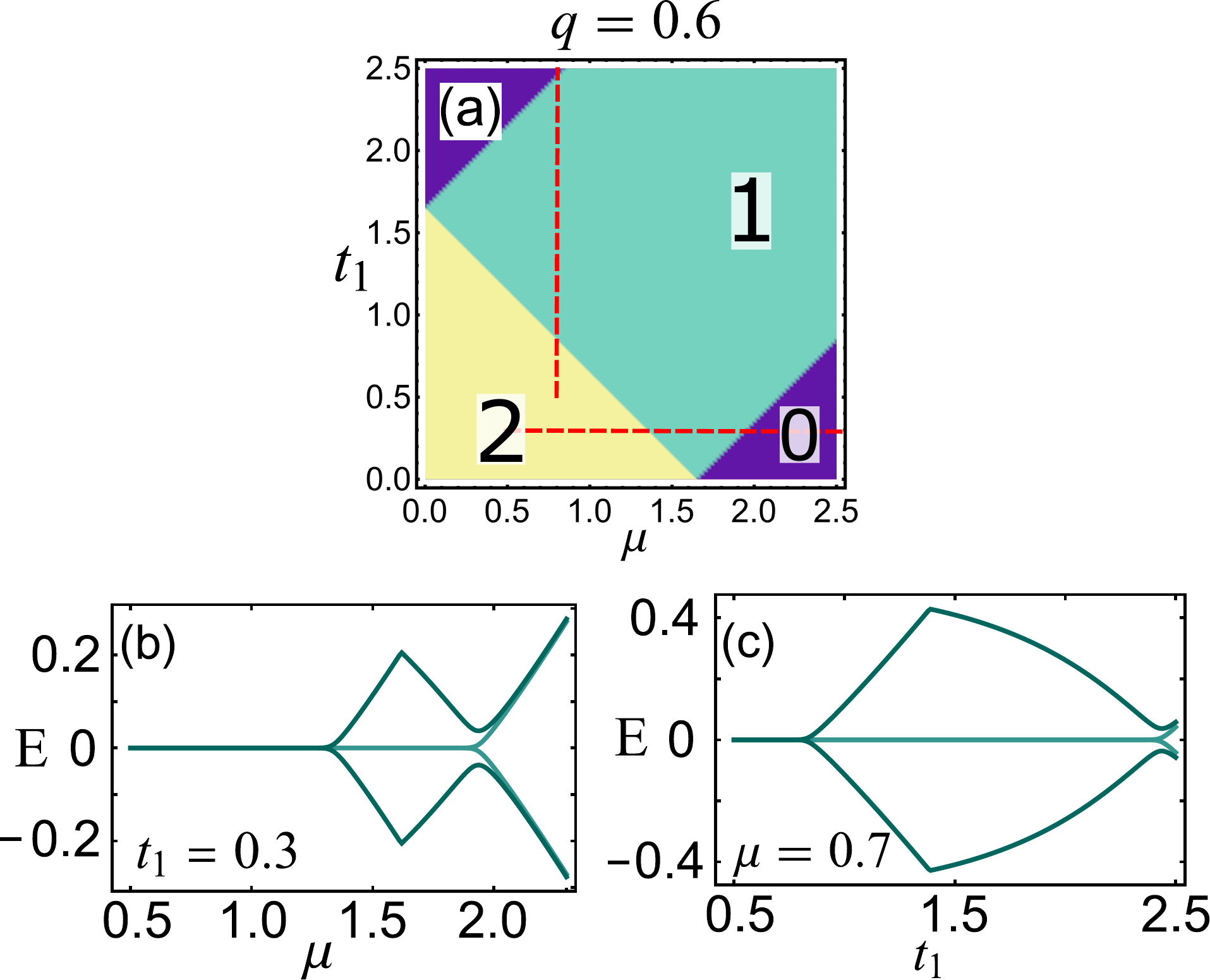}
		\caption{Panel (a) shows the phase diagram of Fig.~\ref{Figure2} for $\Delta_1=0$ and $q=0.6$, together with the horizontal and vertical cuts used to compute the spectra displayed in panels (b) and (c). Panel (b) shows the spectrum as a function of $\mu$ at fixed $t_1=0.3$, while panel (c) shows the spectrum as a function of $t_1$ at fixed $\mu=0.7$. In both cases, the four BdG eigenvalues closest to zero energy are plotted for a ladder composed of two chains, each of length $L=100$. The agreement between the gap closings and the phase boundaries confirms the bulk--edge correspondence.}
		\label{Figure4}
	\end{figure}

We now discuss the numerical results, focusing on the interplay between
current-induced symmetry breaking, topological classification, and
bulk--edge correspondence. To highlight the richness of the phase diagram,
we consider an extended range of parameters, which could in principle be
explored more broadly in cold-atom platforms than in conventional
condensed-matter realizations~\cite{Bloch2012,PhysRevLett.112.043001}. Indeed, in cold-atom implementations, the effective p-wave pairing can be engineered using synthetic interactions~\cite{PhysRevLett.106.220402,PhysRevLett.111.173004}, while the finite current considered here can be naturally realized in a ring geometry threaded by a synthetic gauge flux~\cite{Goldman_2014}, which directly produces the Peierls phase entering our Hamiltonian. Throughout the manuscript, all energies are expressed in units of the longitudinal hopping, $t=1$, and we fix $\Delta=0.8$. This representative choice allows us to explore a broad topological phase diagram while remaining consistent with the parameter regime adopted in previous studies of Kitaev ladders and related geometries~\cite{Maiellaro_2019,Maiellaro2018,Maiellaro2021}. Although this ratio is larger than that typically encountered in conventional condensed-matter systems, it is readily accessible in synthetic quantum platforms, such as ultracold atoms, where hopping and pairing strengths can be engineered independently.\\
Fig.~\ref{Figure0} summarizes the phase diagram of the Kitaev ladder in the presence of finite interchain pairing $\Delta_1 = 0.5$, comparing the equilibrium case ($q=0$) with the current-carrying regime ($q=0.6$). In the absence of current (panel (a)), the system supports multiple gapped topological sectors, including regions characterized by two MBSs per edge, one MBS, and trivial phases, consistently with the underlying $\mathbb{Z}$ classification.\\
When a finite current is introduced [Fig.~\ref{Figure0}(b)], the phase diagram, obtained from $\mathcal{N}$ introduced in Sec.~II, is substantially modified. The two-Majorana sector is completely suppressed, leaving the $\mathcal{N}=1$ phase as the only gapped topological region, characterized by a single Majorana mode per edge. The remaining part of the diagram is redistributed between a gapped trivial region, $\mathcal{N}=0$, and the gapless superconducting region $G$, neither of which supports protected Majorana edge modes. A more systematic analysis of the evolution of the phase diagram over a wider range of $q$, with particular emphasis on the emergence and extension of the gapless region, is presented in Appendix~\ref{App0}.\\
The bulk--edge correspondence is verified in panels (c)--(h), where the probability densities of the lowest-energy Bogoliubov--de Gennes (BdG) modes are shown for representative points selected from the phase diagrams. In the BDI regime (panel (a)), two localized edge modes are clearly visible (panel (c)), while panel (e) corresponds to a single MBS and panel (g) to a bulk state. In the current-carrying regime, the two-mode configuration is lifted (panel (d)), and only one localized edge mode survives (panel (f)), while the trivial phase remains essentially unchanged (panel (h)). These results provide a direct real-space confirmation of the phase diagrams.\\
Further insight is obtained in Fig.~\ref{Figure1}, where the four BdG eigenvalues closest to zero energy are shown along the horizontal and vertical cuts highlighted in Fig.~\ref{Figure0}. Panels (a) and (b) correspond to horizontal cuts at $q=0$ and $q=0.6$, respectively, while panels (c) and (d) show the corresponding vertical cuts. At $q=0$, the zero-energy plateaus reproduce the BDI sectors with two, one, and zero MBSs. At $q=0.6$, panel (b) reveals a qualitatively different spectral structure when the cut enters the gapless region $G$. In the gapped trivial sector, the lowest bulk excitations remain separated from zero by a finite energy gap.By contrast, inside $G$ a dense sequence of quasi-zero-energy states develops. These states are the finite-size counterpart of the zero-energy bulk crossings that emerge in the thermodynamic limit. More precisely, the gapless condition is obtained from $H(k)\psi=0$, which admits non-trivial solutions when $\det H(k)=0$ has real solutions for $k$. The corresponding real momenta identify the bulk gapless points. This behavior must be distinguished from the isolated zero-energy level of the gapped topological sector, which corresponds to a genuine Majorana edge mode. The splitting of the doubly degenerate zero-energy states at finite $q$ highlights the reduced robustness of the two-mode phase under a finite current, in agreement with previous studies driven on open systems~\cite{nano9060894,Maiellaro_NH}.\\
We now turn to the case $\Delta_1 = 0$, where the additional exchange symmetry discussed in Sec.~\ref{Sec2} becomes relevant. Throughout Figs.~\ref{Figure2}--\ref{Figure4} we deliberately restrict the analysis to $q\leq0.6$, so as to remain within the gapped regime and isolate the effect of current-induced symmetry breaking on the Majorana sectors. Indeed, for the parameters used there, $t=1$ and $\Delta=0.8$, the corresponding threshold is $q_c=\arcsin(0.8)\simeq0.927$. Fig.~\ref{Figure2}(a), (c), and (e) display the phase diagrams obtained from the momentum-space invariant for increasing values of the current flux $q$. In contrast with the $\Delta_1 \neq 0$ case, regions supporting two Majorana modes persist even at finite $q$, reflecting the effective decoupling of the system into two independent channels.\\
	The same structure is reproduced by the edge--edge quantum conditional mutual information $I_{ee}$~\cite{Maiellaro2023,PhysRevB.106.155407,PhysRevResearch.4.033088,PhysRevB.107.115160}, shown in panels (b), (d), and (f). This quantity is defined as $I_{ee} = S_{AC} + S_{BC} - S_C - S_{ABC}$, where the different terms correspond to the von Neumann entropies of suitable reduced density matrices obtained by partitioning the system into left edge $A$, right edge $B$, and bulk $C$. The last term $S_{ABC}$ is the total ground-state von Neumann entropy that vanishes whenever the ground state is a pure state~\cite{PhysRevResearch.4.033088,PhysRevB.107.115160}. This specific combination removes classical correlations and isolates genuine quantum correlations between the two edges. As a consequence, $I_{ee}$ acts as a nonlocal topological order parameter, taking quantized values proportional to the number of edge modes. In particular it has been shown that $I_{ee}=E^0_{sq} = log 2/2$, i.e., half of the maximal Bell-pair entanglement, at the exact ground-state topological degeneracy point, for a Kitaev chain with open boundary conditions hosting genuine MBSs~\cite{PhysRevResearch.4.033088,PhysRevB.107.115160}.\\
	A clear agreement is observed between the momentum-space invariant and the entanglement-based phase diagrams, establishing a direct correspondence between the two descriptions. In particular, the different topological sectors are unambiguously identified by distinct values of $I_{ee}$, confirming the validity of the entanglement estimator to indicate topological phases of a ladder in presence of a current flux $q$.\\
	In Fig.~\ref{Figure3}, horizontal and vertical cuts of the entanglement phase diagrams are shown. Panel (a) corresponds to a cut as a function of $\mu$, while panel (b) is taken as a function of $t_1$. For $q=0$, well-defined plateaus at $I_{ee}/E^0_{sq}=2,1,0$ are clearly visible, consistently identifying the three topological sectors. However, for increasing values of $q$, these plateaus become progressively less pronounced and deviate from the ideal quantized values.\\
	This behavior is not related to a modification of the topological structure, but rather to finite-size effects. Indeed, the entanglement calculations are performed on ladders composed of two chains of $L=50$ sites, a size limited by computational convenience. In this regime, the overlap between edge states becomes non-negligible, leading to hybridization effects that reduce the quantization of $I_{ee}$.\\
	To clarify this point, in Fig.~\ref{Figure4} we compare the bulk phase diagram at $q=0.6$ with the corresponding real-space energy spectrum, which can be computed for larger system sizes. Panel (a) shows the phase diagram together with the cuts used to extract the spectra, while panels (b) and (c) display the BdG eigenvalues closest to zero for a ladder with $L=100$ sites per chain. A clear agreement is observed between the phase boundaries and the energy gap closures along the cuts, confirming the validity of the bulk description.\\
	
	\section{Discussion and results}
	\label{Sec4}
	In recent years, signatures compatible with MBSs have been reported in several systems, including semiconductor nanowires with strong spin-orbit coupling proximitized by conventional superconductors~\cite{Mourik2012,Das2012}, planar Josephson junctions~\cite{Fornieri2019,Ren2019}, chains of magnetic adatoms~\cite{NadjPerge2014}, and oxide-based superconducting nanostructures~\cite{PhysRevB.107.L201405}. In all these platforms, the experimental characterization of topological phases is intrinsically based on transport measurements, implying the injection of currents and the coupling of the system to external probes. Understanding how these perturbations modify the topological properties of multichannel isolated superconducting systems therefore represents an important issue for the interpretation of experimental observations.\\
	From this perspective, the effective current-driven description adopted in the present work provides a simple framework to investigate how measurement processes may influence topological protection. In particular, our results show that the current does not simply perturb the low-energy spectrum, but can qualitatively modify the topological classification of the system by breaking protecting symmetries. In addition, sufficiently strong current-induced band tilting can close the excitation gap indirectly, producing a gapless superconducting regime that must be separated from the gapped class-D topological sectors. This effect becomes especially relevant in ladder geometries and multichannel superconductors, where higher topological sectors may exist in equilibrium and can be selectively suppressed under current flow.\\
	At the same time, the persistence of the two-mode phase in the symmetry-protected regime with $\Delta_1=0$ highlights how additional discrete symmetries may stabilize topological sectors even in nonequilibrium conditions. This suggests that the interplay between current-induced symmetry breaking and channel symmetries could represent an important ingredient to design robust multichannel topological superconductors.\\
    Although our analysis has been carried out for the minimal two-leg Kitaev ladder, the underlying mechanism by which a finite quasiparticle momentum modifies the
     symmetry class and the corresponding topological classification is expected to
     extend to more general coupled superconducting systems. In particular,
     multi-leg ladders and multi-band Kitaev models provide a natural framework to
     investigate the interplay between current-induced symmetry breaking and the
     richer topological structures that emerge beyond the conventional tenfold
     Altland--Zirnbauer classification. In this respect, our work may serve as a
     starting point for future investigations of current-driven topological phase transitions in more complex superconducting networks~\cite{Kitaev2009,Bzduszek2017,Bzduszek2026}.\\
	More generally, the agreement observed between the bulk topological invariant, the entanglement-based estimator $I_{ee}$, and the low-energy BdG spectra further supports the use of real-space diagnostics to characterize topological phases under realistic conditions, where finite-size effects, coupling to the environment, and measurement-induced perturbations may invalidate the standard momentum-space description.
	
	\begin{acknowledgments}
		A.M. acknowledges Francesco Romeo, Fabrizio Illuminati and Roberta Citro for fruitful discussions and for the stimulating scientific environment in which this work was developed.\\
		A.M. acknowledges funding from the Horizon Europe EIC Pathfinder programme under Grant No.~101115190 (IQARO).
	\end{acknowledgments}
	
	\appendix
\section{Gapped and gapless superconducting regimes of the Kitaev ladder}
\label{App0}
\begin{figure}
    \centering
    \includegraphics[width=0.9\linewidth]{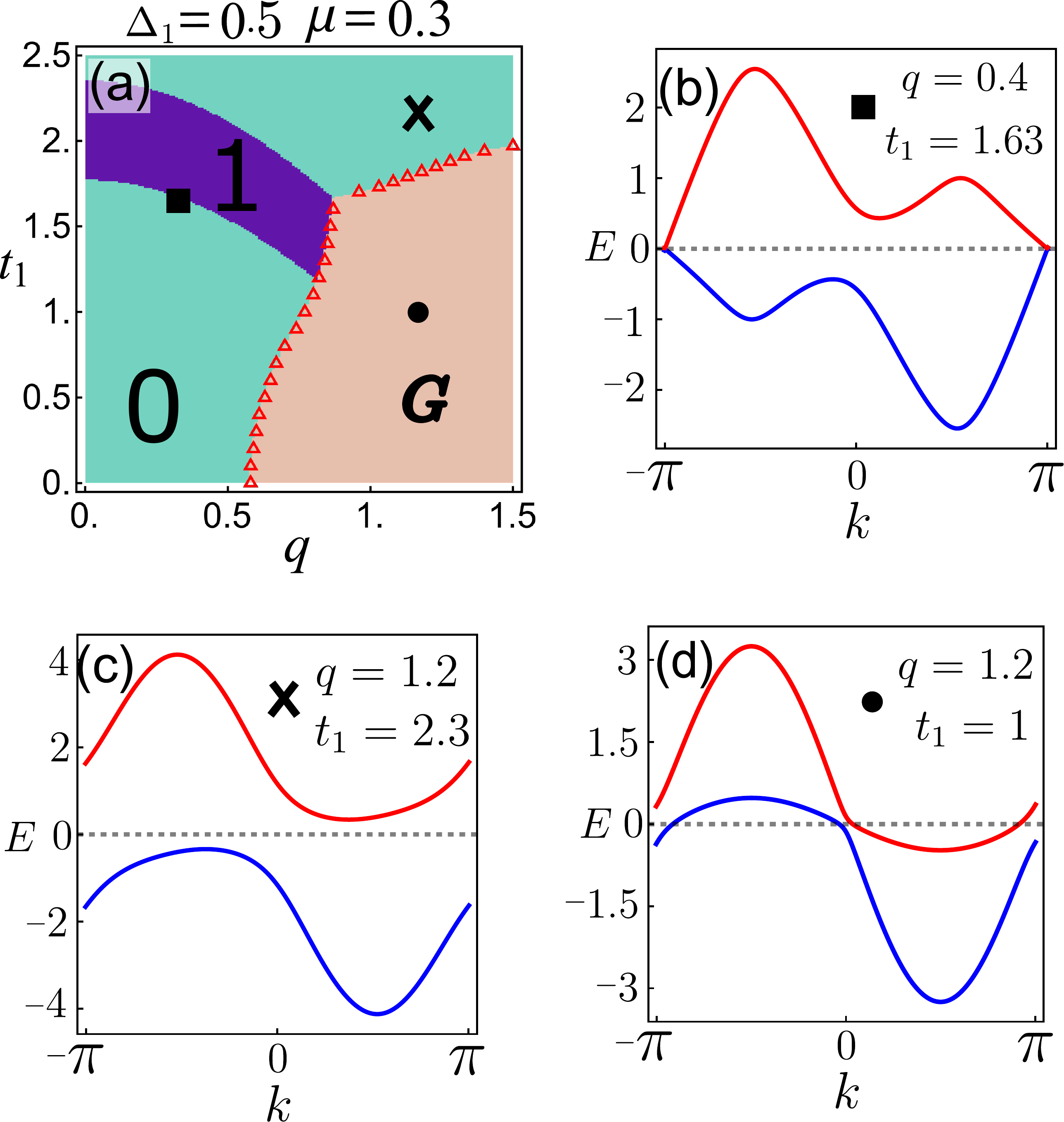}
    \caption{Current-induced gapped and gapless regimes of the Kitaev ladder for $t=1$, $\Delta=0.8$, $\Delta_1=0.5$, and $\mu=0.3$. (a) Phase diagram in the $(q,t_1)$ plane. The labels $0$ and $1$ denote the trivial and topological gapped class-D sectors, respectively, whereas $G$ denotes the gapless superconducting region. The open red triangles mark the onset of the gapless phase, determined from the condition $\det H(k)=0$ as the point where real-momentum solutions first appear. The black symbols indicate the representative parameter points used in panels (b)--(d). Panels (b)--(d) show the two inner BdG branches for the corresponding points: (b) $q=0.4$, $t_1=1.63$, illustrating a gap closing at the particle-hole-invariant momentum; (c) $q=1.2$, $t_1=2.3$, in the gapped topological region; and (d) $q=1.2$, $t_1=1$, in the gapless region, where the upper branch crosses zero at generic momenta and becomes negative over a finite $k$ interval. The dashed horizontal line marks zero energy.}
    \label{FigAppGapless}
\end{figure}

We now make explicit the criterion used in the main text to distinguish the gapped class-D sectors from the current-induced gapless superconducting phase for the general ladder with finite $t_1$ and $\Delta_1$. It is convenient to define
\begin{equation}
\begin{split}
d_0(k)&=-2t\sin k\sin q,\\
\xi(k)&=\mu+2t\cos k\cos q,\\
\delta(k)&=2\Delta\sin k,
\end{split}
\label{eq:generaldefs}
\end{equation}
together with
\begin{equation}
M(k)=t_1^2\left[\xi^2(k)+\Delta_1^2\right]+\Delta_1^2\delta^2(k).
\label{eq:Mgeneral}
\end{equation}
The four eigenvalues of Eq.~\eqref{eq:Hk} can then be written compactly as
\begin{equation}
E_{\alpha,\beta}(k)=d_0(k)+\alpha R_{\beta}(k),
\qquad \alpha,\beta=\pm1,
\label{eq:generalbands}
\end{equation}
\begin{equation}
R_{\beta}(k)=\sqrt{\xi^2(k)+\delta^2(k)+t_1^2+\Delta_1^2+2\beta\sqrt{M(k)}}.
\label{eq:Rgeneral}
\end{equation}
Since $R_+(k)\ge R_-(k)\ge0$, the four branches are ordered as $d_0(k)-R_+(k) \le d_0(k)-R_-(k) \le d_0(k)+R_-(k) \le d_0(k)+R_+(k)$.\\
We therefore identify $d_0(k)\pm R_-(k)$ as the inner branches and define the inner upper branch as 
\begin{equation}
E_+(k)=d_0(k)+R_-(k)
\label{eq:Eplusgeneral}
\end{equation}
In a gapped superconducting phase $E_{+}(k)>0$ for all $k$ (and, by particle--hole symmetry, the corresponding lower branch is negative). The onset of the gapless regime occurs when $E_{+}(k)=0$ at a generic momentum, while inside the gapless phase $E_{+}(k)<0$ over a finite interval of momenta. Equivalently, the current-induced tilt $|d_0(k)|$ exceeds $R_{-}(k)$ for some $k$. This changes the many-body occupation: the upper band becomes occupied for a finite set of momenta and the particle--hole-related lower band becomes empty at the opposite momenta, as in the Lifshitz mechanism discussed in Ref.~\cite{PhysRevResearch.6.033060}.\\
In the decoupled limit $t_1=\Delta_1=0$, Eqs.~\eqref{eq:generalbands}--\eqref{eq:Eplusgeneral} reduce to the current-carrying Kitaev-chain spectrum. In the parameter regime supporting the topological chain, the critical momentum reduces to $q_c=\arcsin(\Delta/t)$, precisely the threshold obtained in Ref.s~\cite{Maiellaro2023,PhysRevResearch.6.033060}. For the coupled ladder, by contrast, $R_{-}(k)$ depends explicitly on $\mu$, $t_1$, and $\Delta_1$, so the boundary of the gapless region must in general be determined from Eq.~\eqref{eq:Eplusgeneral} by minimizing over $k$.\\
Fig.~\ref{FigAppGapless}  provides a direct illustration of the distinction between the gapped topological sectors and the gapless superconducting regime. Panel (a) shows the phase diagram as a function of $q$ and $t_1$ for $\Delta_1=0.5$ and $\mu=0.3$, where the gapless region $G$ is identified by the condition $\min_{k\in(0,\pi)}E_+(k)<0$. Panels (b)--(d) display the corresponding BdG bands for three representative points of the phase diagram. The transition between the two gapped sectors, $\mathcal{N}=0$ and $\mathcal{N}=1$, is associated with a zero-energy gap closing at the particle-hole-invariant momenta, as shown in panel (b). Away from the transition, the gapped phase shown in panel (c) exhibits a finite separation between the positive- and negative-energy branches. By contrast, in the gapless region shown in panel (d), the upper branch crosses below zero energy at generic momentum. These spectra therefore make explicit that the onset of the gapless superconducting regime originates from a finite-$k$ band overlap, which is distinct from the gap closings at $k=0,\pi$ that control the Pfaffian $\mathbb{Z}_2$ invariant.

\section{Gauge-equivalent representation of the current-carrying Hamiltonian}
\label{AppA}
The real-space Hamiltonian introduced in Eq.~(1) is written in a gauge where the supercurrent is encoded through the Peierls phases appearing in the longitudinal hopping,
\begin{equation}
\begin{split}
H
=&
-\sum_{j,l}
\left[
te^{iq}c_{j,l}^{\dagger}c_{j+1,l}
+\Delta c_{j,l}c_{j+1,l}
+\mathrm{h.c.}
\right]
\\
&
+\sum_j
\left[
-t_1c_{j,1}^{\dagger}c_{j,2}
+\Delta_1c_{j,1}c_{j,2}
+\mathrm{h.c.}
\right]
-\mu\sum_{j,l}c_{j,l}^{\dagger}c_{j,l}.
\end{split}
\end{equation}
An equivalent description is obtained by performing the local gauge
transformation
\begin{equation}
c_{j,l}=e^{-iqj}d_{j,l},
\label{GaugeTrans}
\end{equation}
which transfers the Peierls phase from the hopping amplitudes to the pairing
terms.\\
Indeed, the hopping becomes purely real,
\begin{equation}
te^{iq}c_{j,l}^{\dagger}c_{j+1,l}
=
td_{j,l}^{\dagger}d_{j+1,l},
\end{equation}
whereas the intraleg pairing transforms as
\begin{equation}
\Delta c_{j,l}c_{j+1,l}
=
\Delta
e^{-iq(2j+1)}
d_{j,l}d_{j+1,l},
\end{equation}
and the interleg pairing becomes
\begin{equation}
\Delta_1c_{j,1}c_{j,2}
=
\Delta_1
e^{-i2qj}
d_{j,1}d_{j,2}.
\end{equation}
Both pairing channels therefore acquire a spatially varying phase associated
with a finite momentum of the Cooper pairs. After extracting
the site-dependent phase factor, the two pairing amplitudes differ only by an
overall constant phase,
\begin{equation}
\Delta
e^{-iq(2j+1)}
=
\left(\Delta e^{-iq}\right)e^{-i2qj},
\end{equation}
while
\begin{equation}
\Delta_1
e^{-i2qj}
=
\Delta_1e^{-i2qj}.
\end{equation}
This gauge-equivalent representation makes explicit that the current is carried
by Cooper pairs with total momentum $2q$. Importantly, not only the intraleg
pairing $\Delta$, but also the interleg pairing $\Delta_1$, contributes to the
current-carrying condensate. Consequently, a finite supercurrent can be
sustained even in the limit $\Delta=0$, provided that $\Delta_1\neq0$.   
\section{Generalized exchange symmetry in the $t_1=0$ limit}
\label{AppB}
We now consider the additional symmetry-protected limit obtained by setting $t_1=0$ while keeping $\Delta_1\neq0$. In this regime, the two physical egs are coupled only through the inter-leg pairing $\Delta_1$.\\
For $t_1=0$, the Bloch Hamiltonian commutes with a generalized leg-exchange
operator. Up to an overall phase, we choose the representation
\begin{equation}
    \mathcal{M}_y=\sigma_y\otimes\mathbb{I}_2,
\end{equation}
which satisfies
\begin{equation}
    [H(k),\mathcal{M}_y]=0,
    \qquad
    \mathcal{M}_y^2=\mathbb{I}_4.
\end{equation}
Here, $\sigma_y$ acts in the leg subspace. The two eigenvalues of $\mathcal{M}_y$ are therefore $\rho=\pm1$, and the Hamiltonian can be decomposed into the corresponding symmetry sectors.\\
Starting from the Nambu basis
\begin{equation}
    \Psi_k=
    \left(
    c_{k,1},
    c^\dagger_{-k,1},
    c_{k,2},
    c^\dagger_{-k,2}
    \right)^T,
\end{equation}
we introduce the unitary transformation
\begin{equation}
    U_{\rm ex}
    =
    \frac{1}{\sqrt{2}}
    \begin{pmatrix}
        1 & 0 & 1 & 0\\
        0 & 1 & 0 & 1\\
        i & 0 & -i & 0\\
        0 & i & 0 & -i
    \end{pmatrix},
\end{equation}
which diagonalizes the exchange operator. In the transformed basis
$\widetilde{\Psi}_k=U_{\rm ex}^{\dagger}\Psi_k$, the Hamiltonian becomes block diagonal: 
\begin{widetext}
\begin{equation}
\begin{aligned}
    &\widetilde{H}(k)
    =
    U_{\rm ex}^{\dagger}H(k)U_{\rm ex}
    =\\
    &\begin{pmatrix}
        -\mu-2t\cos(k-q)
        &
        -i\left(\Delta_1-2\Delta\sin k\right)
        &
        0 & 0
        \\
        i\left(\Delta_1-2\Delta\sin k\right)
        &
        \mu+2t\cos(k+q)
        &
        0 & 0
        \\
        0 & 0
        &
        -\mu-2t\cos(k-q)
        &
        i\left(\Delta_1+2\Delta\sin k\right)
        \\
        0 & 0
        &
        -i\left(\Delta_1+2\Delta\sin k\right)
        &
        \mu+2t\cos(k+q)
    \end{pmatrix}.
\label{eq:t10_block}
\end{aligned}
\end{equation}
The two $2\times2$ blocks can be written compactly as
\begin{equation}
    H_{\rho}(k)
    =
    \begin{pmatrix}
        -\mu-2t\cos(k-q)
        &
        i \rho \left(\Delta_1+2 \rho \Delta\sin k\right)
        \\
        -i \rho \left(\Delta_1+2 \rho \Delta\sin k\right)
        &
        \mu+2t\cos(k+q)
    \end{pmatrix},
    \qquad
    \rho=\pm1.
\label{eq:Hrho}
\end{equation}
\end{widetext}
Here, $\rho=+1$ and $\rho=-1$ label the two eigenvalues of
$\mathcal{M}_y$.\\
It is useful to introduce
\begin{equation}
\begin{aligned}
d_0(k) &= -2t\sin k\,\sin q,\\
\xi(k) &= \mu + 2t\cos k\,\cos q,\\
m_{\rho}(k) &= -\rho\left(\Delta_1+2\rho\sin k\right).
\end{aligned}
\label{eq:definitions}
\end{equation}
The Hamiltonian of each sector can then be expressed as
\begin{equation}
H_{\rho}(k)
=
d_0(k)\,\sigma_0
-\xi(k)\,\sigma_z
+\Delta_{\rho}(k)\,\sigma_y,
\label{eq:t1zero_pauli}
\end{equation}
where $\sigma_0$ is the identity matrix and $\sigma_{x,y,z}$ are Pauli
matrices.\\
We can also rewrite Eq.~\eqref{eq:Hrho} in a real Majorana basis using the unitary transformation
\begin{equation}
U_M=
\frac{1}{\sqrt{2}}
\begin{pmatrix}
1 & 1\\
i & -i
\end{pmatrix}.
\label{eq:majorana_transformation}
\end{equation}
The transformed block is
\begin{equation}
H_{\rho}^{M}(k)
=
U_M^{\dagger}H_{\rho}(k)U_M
=
\begin{pmatrix}
d_0(k)+\Delta_{\rho}(k) & -\xi(k)\\
-\xi(k) & d_0(k)-\Delta_{\rho}(k)
\end{pmatrix}.
\label{eq:t1zero_majorana_block}
\end{equation}
Therefore, the complete Hamiltonian is the direct sum of two algebraically decoupled effective sectors,
\begin{equation}
H_{t_1=0}^{M}(k)
=
H_{+}^{M}(k)\oplus H_{-}^{M}(k).
\label{eq:t1zero_majorana_full}
\end{equation}
The two sectors have the same normal-state dispersion, while their effective
pairing functions are shifted in opposite directions,
\begin{equation}
\Delta_{\pm}(k)=\Delta_1\pm2\Delta\sin k.
\end{equation}
The corresponding energy bands are
\begin{equation}
E_{\rho,\lambda}(k)
=
d_0(k)
+
\lambda
\sqrt{\xi^2(k)+\Delta_{\rho}^2(k)},
\qquad
\rho,\lambda=\pm1,
\label{eq:t1zero_spectrum}
\end{equation}
or, explicitly,
\begin{widetext}
\begin{equation}
E_{\rho,\lambda}(k)
=
-2t\sin k\sin q
+
\lambda
\sqrt{
\left(\mu+2t\cos k\cos q\right)^2
+
\left(\Delta_1+2\rho \Delta\sin k\right)^2
}.
\end{equation}
\end{widetext}
Because of the term $d_0(k)$ generated by $q$, the two eigenvalues of a fixed sector are not, in general, symmetric around zero at the same momentum. Nevertheless, the complete spectrum satisfies the particle--hole relation
\begin{equation}
E_{\rho,\lambda}(k)
=
-E_{-\rho,-\lambda}(-k).
\label{eq:t1zero_phs_spectrum}
\end{equation}
Thus, a zero-energy state occurring in sector $\rho$ at momentum $k$ is
accompanied by its particle--hole partner in sector $-\rho$ at momentum $-k$.
The particle--hole symmetry therefore exchanges the two blocks, which should
not be regarded as two completely independent class-D superconductors.
A band crosses zero energy whenever
\begin{equation}
E_{\rho,\lambda}(k)=0.
\end{equation}
For $d_0(k)\neq0$, this condition is equivalent to
\begin{equation}
4t^2\sin^2k\,\sin^2q
=
\left(\mu+2t\cos k\cos q\right)^2
+
\left(\Delta_1+2\rho \Delta\sin k\right)^2.
\label{eq:t1zero_generic_closing}
\end{equation}
Equation~\eqref{eq:t1zero_generic_closing} may admit solutions at generic momenta $k\neq0,\pi$, depending on the model parameters. Such solutions occur in particle--hole-related pairs $(\rho,k)$ and $(-\rho,-k)$ and signal the appearance of a gapless spectrum. However, a generic pair of zero-energy crossings at $\pm k$ does not by itself represent a change of the one-dimensional
$\mathbb{Z}_2$ topological invariant.\\
A topological phase transition in a one-dimensional class-D BdG system requires the bulk gap to close at a particle--hole-invariant momentum,
\begin{equation}
k_\star=-k_\star \pmod{2\pi},
\end{equation}
namely
\begin{equation}
k_\star=0,\pi.
\end{equation}
At these momenta one has
\begin{equation}
d_0(k_\star)=0,
\qquad
\Delta_{\rho}(k_\star)=\Delta_1,
\end{equation}
and the spectrum becomes
\begin{equation}
E_{\rho,\lambda}(k_\star)
=
\lambda
\sqrt{
\left(\mu+2t\cos k_\star\cos q\right)^2+\Delta_1^2
}.
\label{eq:t1zero_trim_spectrum}
\end{equation}
In particular,
\begin{equation}
E_{\rho,\lambda}(0)
=
\lambda
\sqrt{
\left(\mu+2t\cos q\right)^2+\Delta_1^2
},
\end{equation}
whereas
\begin{equation}
E_{\rho,\lambda}(\pi)
=
\lambda
\sqrt{
\left(\mu-2t\cos q\right)^2+\Delta_1^2
}.
\end{equation}
Consequently, the gap can close at $k=0$ only if
\begin{equation}
\Delta_1=0,
\qquad
\mu=-2t\cos q,
\label{eq:t1zero_closing_zero}
\end{equation}
while it can close at $k=\pi$ only if
\begin{equation}
\Delta_1=0,
\qquad
\mu=2t\cos q.
\label{eq:t1zero_closing_pi}
\end{equation}
For any finite inter-leg pairing, $\Delta_1\neq0$, the energies at both particle--hole-invariant momenta remain strictly different from zero:
\begin{equation}
\left|E_{\rho,\lambda}(0)\right|\geq|\Delta_1|,
\qquad
\left|E_{\rho,\lambda}(\pi)\right|\geq|\Delta_1|.
\end{equation}
Therefore, $\Delta_1$ acts as a finite mass term at the particle--hole-invariant momenta, removing the gap closings at $k=0$ and $k=\pi$. It does not, in general, prevent an indirect closing at generic $k$. The $t_1=0$, $\Delta_1\neq0$ regime does not support a non-trivial gapped $\mathbb{Z}_2$ topological phase. Depending on the other parameters, the system is either a topologically trivial gapped superconductor or a gapless superconductor characterized by pairs of zero-energy crossings at generic opposite momenta.\\
As a consistency check, let us consider the additional limit $q=0$ . In this case $d_0(k)=0$, and the spectrum becomes
\begin{equation}
E_{\rho,\lambda}(k)=
\lambda
\sqrt{
\left(\mu+2t\cos k\right)^2+
\left(\Delta_1+2 \rho \Delta\sin k\right)^2
}.
\end{equation}
The bulk gap can close only if both terms under the square root simultaneously
vanish, namely
\begin{equation}
\mu+2t\cos k=0,
\qquad
\Delta_1+2 \rho \Delta\sin k=0.
\end{equation}
Using $\sin^2k+\cos^2k=1$, one obtains the transition line
\begin{equation}
\frac{\mu^2}{4t^2}+
\frac{\Delta_1^2}{4\Delta^2}=1.
\label{eq:q0_transition}
\end{equation}
For the parameters adopted in Fig. \ref{Figure0}(a) of the main text, $t=1$, $\Delta=0.8$, and $\Delta_1=0.5$, Eq.~\eqref{eq:q0_transition} gives the critical chemical potential
\begin{equation}
\mu_c
=
2\sqrt{1-\frac{\Delta_1^2}{4\Delta^2}}
\simeq1.90,
\end{equation}
which is in agreement with the phase boundary reported in Fig.~\ref{Figure0}(a) of the main text. This provides an additional consistency check of the analytical treatment presented in this Appendix.
\bibliography{Bib}
\end{document}